%% file: main.tex
\documentclass[conference]{IEEEtran}
\IEEEoverridecommandlockouts
\usepackage{cite}
\usepackage{amsmath,amssymb,amsfonts}
\usepackage{graphicx}
\usepackage{textcomp}
\usepackage{xcolor}
\input{commands}

\def\BibTeX{{\rm B\kern-.05em{\sc i\kern-.025em b}\kern-.08em
    T\kern-.1667em\lower.7ex\hbox{E}\kern-.125emX}}
\usepackage{algpseudocode}  
\usepackage{algorithm}  
\usepackage{url}
\usepackage{comment}

\usepackage{dirtytalk}
\begin{document}

\title{Admission Control in 5G Networks for the Coexistence of eMBB-URLLC Users
\thanks{Authors would like to certify that this work has not been published in any other conference or has not been submitted for any other publication elsewhere.}}

\author{\IEEEauthorblockN{Nipuni Uthpala Ginige, K. B. Shashika Manosha, Nandana Rajatheva, and Matti Latva-aho}
\IEEEauthorblockA{Center for Wireless Communications,
University of Oulu,
Finland \\
\{nipuni.ginige, manosha.kapuruhamybadalge, nandana.rajatheva, matti.latva-aho\}@oulu.fi}

}

\maketitle

\begin{abstract}
In this paper, we consider the problem of admission control in 5G networks where enhanced mobile broadband (eMBB) users and ultra-reliable low-latency communication (URLLC) users are coexisting. URLLC users require low latency and high reliability while eMBB users require high data rates. Thus, it is essential to control the admission of eMBB users while giving priority to all URLLC users in a network where both types of users are coexisting. Our aim is to maximize the number of admitted eMBB users to the system with a guaranteed data rate while allocating resources to all URLLC users. 
We formulated this as an ~$\ell_0$ minimization problem. Since it is an NP-hard problem we have used  approximation methods and sequential convex programming to obtain a suboptimal solution. Numerically we have shown that the proposed algorithm achieves near-optimal performance. Our algorithm is able to maximize the number of admitted eMBB users with an optimal allocation of resources while giving priority to all URLLC users.
\end{abstract}


\section{Introduction}

Fifth generation (5G) new radio (NR) supports three main use cases. They are enhanced mobile broadband (eMBB), ultra-reliable low-latency communications (URLLC) and massive machine-type communications (mMTC). According to 3GPP, quality of service (QoS) requirements of URLLC is ultra-high reliability and low transmission latency, whereas eMBB requires high data rates \cite{8718159}. The coexistence of eMBB and URLLC users in the same resource is a difficult issue since simultaneously achieving high data rates for eMBB users and the ultra reliability and low latency for URLLC users becomes a challenging scheduling task due to the trade-off between latency, reliability and achieving high data rates.  

Admission control in wireless networks can be interpreted as finding the maximum amount of traffic or maximum number of users that can be admitted simultaneously to the system while efficiently using the available resources and satisfying QoS requirements. The admission control problem is formulated as an~$\ell_0$ minimization problem in \cite{8241816} for multi-cell downlink multiple-input single-output (MISO) system. Authors in \cite{4570234} have proposed  two algorithms to solve the optimization problem of admission control using semi-definite relaxation method and the second-order cone programming method. Multi-user admission control and beamformer optimization for the MISO heterogeneous networks is considered in \cite{4570234}. Authors in \cite{5682069} are suggesting suboptimal greedy search algorithm for solving the admission control problem and finding optimal power and bandwidth allocations. 

Effective bandwidth is the minimum amount of the bandwidth required to satisfy QoS requirements \cite{7945856}.  If the maximum achievable rate of the URLLC user is greater than or equal to the effective bandwidth, which was derived using the reliability and latency values, we can say that reliability and latency requirement of the URLLC user is guaranteed \cite{7945856, 8070468, 8541123, 8269133}. Most of the scheduling algorithms for the coexistence of eMBB and URLLC in literature suggest puncturing eMBB users, in order to give priority to URLLC users and satisfy their reliability and latency requirements \cite{8718159,  8408793 , 8746407, 8761648}. 

URLLC users are time and mission critical, therefore they need to be given priority when they have something to transmit. However, eMBB users are best effort users, they need higher data rate than the other requirements. Thus, to accommodate it we need to control the admissions of eMBB users and facilitate scheduling all the URLLC users. Therefore, it is essential to know the possible number of eMBB users that can be supported by the system while meeting the reliability and latency requirements of URLLC users. However, no research has been found which considers the admission control problem in the wireless network where eMBB users and URLLC users are coexisting. 

In this paper, we propose an algorithm to solve the admission control problem in 5G networks where eMBB and URLLC users are coexisting\cite{thesis} \footnote{This paper is based on the research findings of the first author's master's thesis \cite{thesis}}, \cite{2019admission} \footnote{A pre-conference version of this papaer is published in \cite{2019admission}}. The objective of the problem is to maximize the number of admitted eMBB users under four constraints: 1) signal-to-interference-plus-noise ratio constraint for eMBB users which is derived through Shannon's rate, 2) signal-to-noise ratio constraints for URLLC users in order to satisfy high reliability and low latency requirements of URLLC users which is derived through the approximation of Shannon's rate in short blocklength regime and simplified using the notion of effective bandwidth to obtain a lower bound, 3) transmit power constraint and 4) total bandwidth constraint. 

The proposed algorithm finds the maximum number of eMBB users who have sufficient data rates that can be admitted to the system while allocating power, bandwidth and beamforming directions to all URLLC users whose latency and reliability requirements are always guaranteed. The algorithm is derived using sequential convex programming. Numerically, we show that the proposed algorithm achieves near-optimal performance.

The rest of the paper is organized as follows. The system model and the problem formulation are presented in Section \ref{system model}. The proposed admission control algorithm is presented in Section \ref{algo}. The numerical results are presented in Section \ref{results} and Section \ref{conclusion} concludes our paper.


\section{System model and Problem Formulation}
\label{system model}
\begin{figure}[ht]
	\centering
	\includegraphics[width=0.4\textwidth]{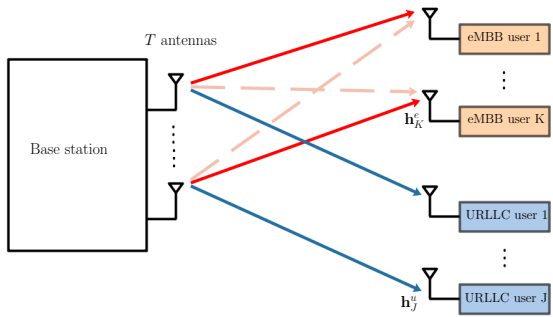}
	\caption{Illustration of the system model.}
	\vspace{-2mm}
	\label{fig:Illustration of the System Model}
\end{figure}

We consider the downlink of a single-cell MISO system where eMBB users and URLLC users are coexisting as shown in Fig.\ref{fig:Illustration of the System Model}.  We assume that the base station have $T$ transmit antennas. The set of all the users in the network is denoted by $\mathcal U$. The set of all eMBB users denoted by {$\mathcal U_e$} $\subset$ {$\mathcal U$} and they are labeled with the integer values $k=1, \ldots, K$. We use the notation $\mathcal U_u$ $\subset$ {$\mathcal U$} to denote the set of all URLLC users and they are labeled with the integer values $j=1, \ldots, J$. We assume that all users have only one receive antenna. 

We consider eMBB and URLLC users coexist using orthogonal spectrum sharing approach \cite{8761648}. Let the total bandwidth of the system be $B_{total}$, total bandwidth for eMBB users be $b^e$ and that of URLLC users be $b^u$. Note that there is no interference between eMBB users and URLLC users, since they are getting a separate portion of bandwidth. 
Furthermore, we consider orthogonal frequency division multiple access (OFDMA) for URLLC users. Thus, the URLLC users are scheduled in orthogonal resources, hence there is no interference between URLLC users.

We assume that bandwidth allocation for all eMBB users and for each URLLC user is less than or equal to the total bandwidth of the system $B_{total}$, i.e., 

\begin{equation}
\label{bandwidth constraint}
b^e + {\sum}_{j=1 }^{J}b_j^u \leq B_{total},
\end{equation} 
where  $b_{j}^u$ is the bandwidth allocated to $j$th URLLC user.

The signal vector transmitted by the base station is given by
${\mathbf{x}} =  \sum_{k\in \mathcal U} \mathbf{m}_kd_k$,
where $d_{k}$ is the normalized data symbol of $k$th user, and  we assume that the data streams are independent. The beamforming vector of $k$th user, $\mathbf{m}_k\in\mathbb{C}\tran$ can be written as $\mathbf{m}_k = \sqrt{p_k} \mathbf{u}_k$,  where $\mathbf {u}_k \in \mathbb C\tran$ is the normalized beamformer and $p_k$ is the power of $k$th user. 

Received signal vector of $k$th eMBB user is given by

\begin{equation}
 \label{Received signal for eMBB user}
y_{k}^e=(\mathbf{h}_{k}^{e})\herm \mathbf{m}_{k}^e d_{k}^e + {\sum}_{i=1 ,i\neq k}^{K} (\mathbf{h}_{k}^e)\herm \mathbf{m}_{i}^e d_{i}^e+w_{k}^e,
\end{equation}
where $\mathbf{h}_{k}^{e} \in \mathbb C\tran$ is the channel vector from base station to eMBB user $k$,  ${\bf{m}}_{k}^e\in \mathbb C\tran$ is the beamforming vector of $k$th eMBB user  and $w_{k}^e \sim CN(0,\sigma^2_e)$  is the additive white Gaussian noise (AWGN) at eMBB user $k$. We consider the noise variance as $\sigma^2_e=N_0b^e$, where $N_0$ is single-sided noise spectral density. 

Received signal of $j$th URLLC user can be written as

 \begin{equation}
 \label{Received signal for URLLC user}
 y_{j}^u=(\mathbf{h}_{j}^u)\herm \mathbf{m}_{j}^u d_{j}^u +w_{j}^u,
 \end{equation}
where $\mathbf{h}_{j}^{u} \in \mathbb C\tran$ is the channel vector from base station to URLLC user $j$,  ${\bf{m}}_{j}^u\in \mathbb C\tran$ is the beamforming vector of $j$th URLLC user and $w_{j}^u \sim CN(0,\sigma_{j,u}^2)$  is the additive white Gaussian noise (AWGN) at URLLC user $j$. We consider the noise variance as $\sigma_{j,u}^2=N_0b_{j}^u$. 
 
The received SINR of $k$th eMBB user can be expressed as

\begin{equation}
\label{SINR for eMBB user}
\gamma_{k}^e =\frac{\vert (\mathbf{h}_{k}^e)\herm \mathbf{m}_{k}^e \vert^{2} }{{\sum}_{i=1 ,i\neq k}^{K}  \vert (\mathbf{h}_{k}^e)\herm \mathbf{m}_{i}^e\vert^{2} +N_0b^e}.
\end{equation}

The received SINR of $j$th URLLC user can be expressed as

\begin{equation}
\label{SINR for URLLC user}
\gamma_{j}^u = \frac{\vert (\mathbf{h}_{j}^u)\herm\mathbf{m}_{j}^u\vert ^{2}}{N_0b^u_j}.
\end{equation}

The maximum achievable rate for $k$th eMBB user can be written as 

\begin{equation}
\label{Data rate for eMBB user}
R_{k}^e=b^e\log_{2}(1+{\gamma }_{k}^e).
\end{equation}

We assume that the target rate for an eMBB user is $R_{target}$. Thus, the target SINR for $k$th eMBB user, $\gamma_k^{e,tar} $ can be expressed as

\begin{equation}
\label{SINR target for eMBB user k}
\gamma_k^{e,tar} = 2^{\frac{R_{target}}{b_e}} -1.
\end{equation}

The target rate for eMBB users can be achieved if its SINR, which is mentioned in \eqref{SINR for eMBB user}, is greater than the SINR threshold, $\gamma_k^{e,tar} $, i.e., 

\begin{equation}
\label{SINRconstraint for embb}
\gamma_k^e \geq \gamma_k^{e,tar}.
\end{equation}

The overall packet loss probability requirement of an URLLC user is the probability of the complement of overall reliability requirement. 
The overall packet loss probability, $\epsilon$ can be expressed as $\epsilon = \epsilon_c+ \epsilon_q$, where $\epsilon_c$  is the transmission-error probability and $\epsilon_q$  is the queuing-delay violation probability.  

Furthermore, we assume that downlink transmission only requires one frame and duration of one frame is $T_f$. Moreover, the latency of the backhaul is equal to the duration of one frame, $T_f$. Thus, we can obtain end to end queuing delay by $D_q = D_{max}- 2T_{f}$, where $D_{max}$ is the maximum packet delay threshold. 

If channel state information (CSI) is known at the transmitter and receiver, in quasi-static, interference-free, flat fading channel, the achievable rate of $j$th user can be approximated by
\cite{8269133},

\begin{equation}
\label{Data rate for urllc users}
R_{j}^u= \frac{\tau b_{j}^u}{\ln2}[\ln(1+{ \gamma }_{j}^u)- \sqrt{\frac{V_{j}^u}{\tau b_{j}^u}}Q^{-1}(\epsilon_c)] \hspace{2mm}bits/frame,
\end{equation}
where $\tau$ is the duration for data transmission in one frame,  $Q^{-1}$ is the inverse Q function and  $V_{j}^u$ is channel dispersion of URLLC user $j$, which is given by
\begin{equation}
\label{channel dispersion}
V_{j}^u = 1 - \frac{1}{(1+{ \gamma }_{j}^u)^2}.
\end{equation}

The channel coherence time is greater than the end to end delay since URLLC has an end to end delay less than 1 ms. This means URLLC users have a quasi-static channel and the rate of URLLC users, which is mentioned in \eqref{Data rate for urllc users}, can consider as a constant for a given resource allocation policy. The queuing delay requirements ($D_q$ and $\epsilon_q$) can be satisfied when this constant achievable rate is greater than or equal to the effective bandwidth \cite{8070468 ,8269133, 8541123, 7945856}. The effective bandwidth for a Poisson process with arrival packet rate $\lambda$, can be expressed as \cite{8070468},

\begin{equation}
\label{effective bandwidth}
E^B = \frac{\mu T_{f} \ln{\frac{1}{\epsilon_q}}}{D_{q}\ln{(\frac{T_{f}\ln{\frac{1}{\epsilon_q}}}{\lambda D_{q}}+1)}} \hspace{2mm}bits/frame,
\end{equation}
where $\mu$ is the number of bits contained in each packet. We can obtain a lower bound for the SNR required to satisfy queuing delay requirements by substituting $ R_{j}^u=E^B$ and $V_{j}^u \approx 1$. Thus, the threshold for SNR of URLLC user $j$ is given by

\begin{equation}
\label{snr required}
\gamma_{j}^{u,tar}= \exp{[\frac{E^B\ln2}{\tau b_{j}^u}+\sqrt{\frac{1}{\tau b_{j}^u}}Q^{-1}(\epsilon_c)]}-1.
\end{equation}

Latency and reliability requirements of $j$th URLLC user is satisfied if SNR of $j$th URLLC user, which is mentioned in \eqref{SINR for URLLC user}, is greater than the SINR threshold $\gamma_{j}^{u,tar}$, i.e.,

\begin{equation}
\gamma_j^u \geq \gamma_j^{u,tar}.
\end{equation}

We assume that the total power allocated for both eMBB and URLLC users is less than or equal to maximum transmit power at the base station $P_{total}$, i.e., 

\begin{equation}
\label{sum power constraint}
{\sum}_{k=1 }^{K} \parallel \mathbf{m}_{k}^e \parallel_2^2 + {\sum}_{j=1 }^{J} \parallel\mathbf{m}_{j}^u \parallel_2^2 \leq P_{total}.
\end{equation}

It is needed to prioritize URLLC users due to their low latency and high reliability requirements. However, eMBB users require high data rates. We have to control the admission of eMBB users in order for their coexistence.
Hence, our goal is to maximize the number of admitted eMBB users such that all the constraints related to eMBB and URLLC users are satisfied. Thus, we need to maximize the cardinality of $\mathcal U_e$. To formulate this problem as a mathematical optimization problem we define the non negative auxiliary variable $s_{k}$ and relax the SINR constraint for $k$th eMBB user as follows:

\begin{equation}
\label{relax SINRconstraint for embb}
\gamma_k^e \geq \gamma_k^{e,tar} - s_{k}.
\end{equation}

In  \eqref{relax SINRconstraint for embb}, we can obtain \eqref{SINRconstraint for embb} when $s_{k}=0$. That means when $s_{k}=0$ the SINR constraint of $k$th eMBB user is satisfied. Therefore, in order to maximize the number of admitted eMBB users who achieve the target rate, we have to minimize the number of users that require a strictly positive value of auxiliary variable $s_k$. In other words we have to make maximum number of $s_{k}$'s to be zeros. Hence the problem of admission control for eMBB in the coexistence of URLLC and eMBB users can be expressed as

\begin{subequations}
	\label{Optproblem1}
	\begin{align}
	\mbox{minimize } &\quad \textstyle  {\parallel {\mathbf s} \parallel}_0 \nonumber\\
	\mbox{subject to} &\quad  \gamma_k^e \geq \gamma_k^{e,tar}-s_k,\forall \hspace{2mm} k \in \mathcal U_e \label{Optproblem_Con1.1}\\
	&\quad \gamma_{j}^u \geq \gamma_{j}^{u,tar} ,\forall \hspace{2mm} j \in \mathcal U_u  \label{Optproblem_Con1.2}\\
	&\quad  b^e + {\sum}_{j=1 }^{J}b_j^u \leq B_{total}
	\label{Optproblem_Con1.3}  \\	&\quad b^e \geq 0\label{Optproblem_Con1.8}\\
	&\quad b_{j}^u \geq 0,\forall \hspace{2mm} j \in \mathcal U_u	\label{Optproblem_Con1.5}\\&\quad {\sum}_{k=1 }^{K} \parallel \mathbf{m}_{k}^e \parallel_2^2 + {\sum}_{j=1 }^{J} \parallel \mathbf{m}_{j}^u \parallel_2^2 \leq P_{total}\label{Optproblem_Con1.6} \\	&\quad s_{k} \geq 0,\forall \hspace{2mm} k \in \mathcal U_e,	\label{Optproblem_Con1.7}
	\end{align}
\end{subequations}
where $\mathbf s = [s_1,.....,s_k]\tran$ and optimization variables are \{$s_{k}$, $\mathbf{m}_{k}^e$\} $\forall \hspace{2mm} k \in \mathcal U_e$, $b^e$ and \{$\mathbf{m}_{j}^u$, $b_{j}^u$\} $\forall \hspace{2mm} j \in \mathcal U_u$.

\section{Algorithm derivation}
\label{algo}
Problem \eqref{Optproblem1} has an $\ell_0$ objective function and it is known as an NP-hard problem. Thus, it is exponentially complex to find an optimal solution to this problem. We provide a suboptimal algorithm that can find a suboptimal solution to the problem. The proposed algorithm is based on $\ell_0$ approximation method and sequential convex programming.

We approximate the objective function with a concave function ${\sum}_{k=1 }^{K} \log (s_k + \delta)$ where $\delta$ is small positive constant and $s_k \geq 0$, $\forall \hspace{2mm} k \in \mathcal U_e$\cite{candes2008enhancing}. We denote the interference plus noise experienced by $k$th eMBB user, by the variable $\beta_k$ as $\beta_k = {{\sum}_{i=1 ,i\neq k}^{K}  \vert (\mathbf{h}_{k}^e)\herm \mathbf{m}_{i}^e\vert^{2} +N_0b^e}$. Hence, the original optimization problem \eqref{Optproblem1} can be approximated as the following optimization problem:

\begin{subequations}
	\label{Optproblem2}
	\begin{align}
	\mbox{minimize } &\quad \textstyle {\sum}_{k=1 }^{K} \log(s_k + \delta) \nonumber\\
	\mbox{subject to} &\quad    2^{\frac{R_{target}}{b^e}} -1-s_{k} - \frac{\vert (\mathbf{h}_{k}^e)\herm \mathbf{m}_{k}^e \vert^{2} }{\beta_k}\leq 0,\forall \hspace{2mm} k \in \mathcal U_e \label{Optproblem_Con2.1}\\&\quad {{\sum}_{i=1 ,i\neq k}^{K}  \vert (\mathbf{h}_{k}^e)\herm \mathbf{m}_{i}^e\vert^{2} +N_0b^e} \leq \beta_k, \forall \hspace{2mm} k \in \mathcal U_e \label{Optproblem_Con2.2} \\ &\quad	 \exp{[\frac{E^B\ln2}{\tau b_{j}^u}+\sqrt{\frac{1}{\tau b_{j}^u}}Q^{-1}(\epsilon_c)]}-1 \nonumber\\
	& \hspace{15.5mm} -\frac{{\vert (\mathbf{h}_{j}^u)\herm\mathbf{m}_{j}^u\vert ^{2}}}{N_0b^u_j}\leq 0 ,\forall \hspace{2mm} j \in \mathcal U_u  \label{Optproblem_Con2.3}\\
	&\quad \mbox{constraints~}  \eqref{Optproblem_Con1.3}, \eqref{Optproblem_Con1.8}, \eqref{Optproblem_Con1.5}, \eqref{Optproblem_Con1.6}, \eqref{Optproblem_Con1.7},  \label{Optproblem_Con2.4}
	\end{align}
\end{subequations}
where optimization variables are \{$s_{k}$, $\mathbf{m}_{k}^e$\} $\forall \hspace{2mm} k \in \mathcal U_e$, $b^e$ and $\{\mathbf{m}_{j}^u$, $b_{j}^u\}$ $\forall \hspace{2mm} j \in \mathcal U_u$.

Problem \eqref{Optproblem2} is  still non convex, because still it has a concave objective function and non convex constraint functions, i.e., the constraints \eqref{Optproblem_Con2.1} and \eqref{Optproblem_Con2.3} are non convex. Therefore, to solve the the problem \eqref{Optproblem2}, we apply sequential convex programming \cite{scp}.  

We denote the objective function of the problem \eqref{Optproblem2}  by $f(s) = {\sum}_{k=1 }^{K} \log(s_k + \delta)  $. Since $f(s)$ is a concave function \cite[Ch.~3]{boyd2004convex}, we consider its first order approximation, and approximation of the objective function can be denoted by \cite{scp},
\begin{equation}
\label{eq22}
\hat{f}({\mathbf s}) = f(\hat{\mathbf s}) +  {\sum}_{k=1 }^{K}{(s_{k} - \hat s_{k})}/{(\hat s_{k} + \delta)},
\end{equation}
and it is evaluated at the point $\hat{\mathbf s}= [\hat s_1,...., \hat s_K]$.

The constraint \eqref{Optproblem_Con2.1} is in the form of `difference of convex' function. We apply convex - concave procedure to make the constraint  \eqref{Optproblem_Con2.1} convex \cite{scp}.  We define $g_k(\mathbf{m}_k^e,\beta_k)$ as  $g_k(\mathbf{m}_k^e,\beta_k)= {\vert (\mathbf{h}_{k}^e)\herm \mathbf{m}_{k}^e \vert^{2} }/{\beta_k} $. First order approximation of $g_k(\mathbf{m}_k^e,\beta_k)$  is as follows:
\begin{multline}
  \label{eq23}
\hat{g}_{k}({\mathbf {m}}_{k}^e,\beta_{k}) =  g_{k}(\hat{\mathbf {m}}_{k}^e,\hat {\beta}_{k}) \\ + {\nabla g_{k}(\hat{\mathbf {m}}_{k}^e,\hat{\beta}_{k})}\tran (({\mathbf {m}}_{k}^e,\beta_{k})-(\hat{\mathbf {m}}_{k}^e,\hat{\beta_{k}})),  
\end{multline}
where $\nabla g_{k}(\hat{\mathbf {m}}_{k}^e,\hat{\beta}_{k})$ is the gradient of $\textstyle g_{k}({\mathbf {m}}_{k}^e,\beta_{k})$ which is evaluated at the point $(\hat{\mathbf {m}}_{k}^e,\hat{\beta}_{k})$. $\nabla g_{k}(\hat{\mathbf {m}}_{k}^e,\hat{\beta}_{k})$ is given by

\begin{equation}
\label{eq24}
\displaystyle  \nabla g_{k}(\hat{\mathbf {m}}_{k}^e,\hat{\beta}_{k}) =(\frac{2 {\mathbf {h}}_k^e ({\mathbf {h}}_k^e)\herm \hat{\mathbf m}_{k}^e}{\hat\beta_{k}}, \frac{-(\hat{\mathbf m}_{k}^e)\herm {\mathbf {h}}_{k}^e ({\mathbf {h}}_{k}^e)\herm\hat{\mathbf m}_{k}^e} {\hat\beta_{k}^2}).
\end{equation}

The constraint \eqref{Optproblem_Con2.3} is also in the form of `difference of convex' function. We apply convex - concave procedure to make the constraint \eqref{Optproblem_Con2.3} convex \cite{scp}. We define $z_j(\mathbf{m}_j^u,b^u_j)$ as,  $z_j(\mathbf{m}_j^u,b_j^u)={{\vert (\mathbf{h}_{j}^u)\herm \mathbf{m}_{j}^u \vert^{2} }}/{N_0b_j^u}$. The first order approximation of  $z_j(\mathbf{m}_j^u,b^u_j)$ is as follows:

\begin{multline}
\label{eq25}
\hat{z}_{j}({\mathbf {m}}_{j}^u,b_j^u) =  z_{j}(\hat{\mathbf {m}}_{j}^u,b_j^u) \\ + {\nabla z_{j}(\hat{\mathbf {m}}_{j}^u,\hat b_j^u)}\tran (({\mathbf {m}}_{j}^u,b_j^u)-(\hat{\mathbf {m}}_{j}^u,\hat b_j^u)),
\end{multline}
where $\nabla z_{j}(\hat{\mathbf {m}}_{j}^u,\hat b_j^u)$ is the gradient of $\textstyle z_{j}({\mathbf {m}}_{j}^u, b_j^u)$ which is evaluated at the point~$(\hat{\mathbf {m}}_{j}^u,\hat b_j^u)$. $\nabla z_{j}(\hat{\mathbf {m}}_{j}^u,\hat b_j^u)$ is given by

\begin{equation}
\label{eq26}
\displaystyle  \nabla z_{j}(\hat{\mathbf {m}}_{j}^u,\hat b_j^u) =(\frac{2 {\mathbf {h}}_j^u ({\mathbf {h}}_j^u)\herm \hat{\mathbf m}_{j}^u}{N_0\hat {b_{j}^u}}, \frac{-(\hat{\mathbf m}_{j}^u)\herm {\mathbf {h}}_{j}^u ({\mathbf {h}}_{j}^u)\herm\hat{\mathbf m}_{j}^u} {N_0{(\hat{b}_{j}^{u})}^2}).
\end{equation}

Now by using expressions \eqref{eq22}, \eqref{eq23} and \eqref{eq25}, we approximate the problem \eqref{Optproblem2} as the following convex optimization problem:

\begin{subequations}
	\label{Optproblem3}
	\begin{align}
	\mbox{minimize } &\quad \textstyle {\sum}_{k=1 }^{K} {s_{k}}/{(\hat{s}_{k} + \delta)}  \nonumber\\
	\mbox{subject to} &\quad    2^{\frac{R_{target}}{b^e}} -1-s_{k} - \hat{g}_{k}({\mathbf {m}}_{k}^e,\beta_{k}) \leq 0,\forall \hspace{2mm} k \in \mathcal U_e \label{Optproblem_Con3.1}\\&\quad {{\sum}_{i=1 ,i\neq k}^{K}  \vert (\mathbf{h}_{k}^e)\herm \mathbf{m}_{i}^e\vert^{2} +N_0b^e} \leq \beta_k, \forall \hspace{2mm} k \in \mathcal U_e \label{Optproblem_Con3.2} \\ &\quad	  \exp{[\frac{E^B\ln2}{\tau b_{j}^u}+\sqrt{\frac{1}{\tau b_{j}^u}}Q^{-1}(\epsilon_c)]}-1 \nonumber\\
	& \hspace{19.5mm}- \hat{z}_{j}({\mathbf {m}}_{j}^u,b_j^u)\leq 0 ,\forall \hspace{2mm} j \in \mathcal U_u \label{Optproblem_Con3.3}\\
	&\quad \mbox{constraints~}   \eqref{Optproblem_Con1.3}, \eqref{Optproblem_Con1.8}, \eqref{Optproblem_Con1.5}, \eqref{Optproblem_Con1.6}, \eqref{Optproblem_Con1.7},	\label{Optproblem_Con3.4}
	\end{align}
\end{subequations}
where the optimization variables are \{$s_{k}$, $\mathbf{m}_{k}^e$\} $\forall \hspace{2mm} k \in \mathcal U_e$, $b^e$ and $\{\mathbf{m}_{j}^u$, $b_{j}^u \}$ for $\forall \hspace{2mm} j \in \mathcal U_u$. We have dropped the constant terms $f(\hat {\mathbf s})$ and ${\hat s_k}/{\hat s_k + \delta}$ from the objective function of problem \eqref{Optproblem3}, since they are not affecting the solution.

The proposed algorithm for solving problem \eqref{Optproblem3} is summarized in \emph{Algorithm~\ref{algorithm 1}}.

\begin{algorithm}
	\caption{Algorithm for solving problem \eqref{Optproblem3}}\label{algorithm 1}
	\begin{algorithmic}[1]
		
		\State  {\bf{initialization}}: $ \{s_{k}^0$, $({\mathbf {m}}_{k}^{e})^0$, $\beta_{k}^0\}$ $\forall \hspace{2mm} k \in \mathcal U_e, $ $b^{e}$ and \{$(\mathbf{m}_{j}^{u})^0$, $(b_{j}^{u})^0$\} $\forall \hspace{2mm} j \in \mathcal U_u$, iteration index~$p=0$.
		\label{step_initialization}  
		\Statex \textbf{repeat}
		\State Set $\hat {\mathbf{m}}_{k}^{e} = (\mathbf{m}_{k}^{e})^{p}$, $\hat{\beta}_{k}  =\beta_{k}^{p}$ $\forall \hspace{2mm} k \in \mathcal U_e$ and  $\hat {\mathbf{m}}_{j}^u = (\mathbf{m}_{j}^{u})^{p}$, $\hat {{b}}_{j}^u = ({b}_{j}^{u})^{p}$  $\forall \hspace{2mm} j \in \mathcal U_u$. Form $\hat{g}_{k}({\mathbf {m}}_{k}^e,\beta_{k}) \hspace{2mm} \forall k$ using \eqref{eq23} and $\hat{z}_{j}({\mathbf {m}}_{j}^u,b_j^u) \hspace{2mm} \forall j$ using \eqref{eq25}.
		 \label{step_2}
		\State Solve problem \eqref{Optproblem3}. Denote the solution $\{s_{k}^{\star}$, ${({\mathbf {m}}_{k}^{e})}^\star$, $\beta_{k}^{\star}\}$ $\forall \hspace{2mm} k \in \mathcal U_e$ and $\{({\mathbf {{m}}_{j}^{u})}^\star$, ${(b_{j}^{u})}^\star \}$ $\forall \hspace{2mm} j \in \mathcal U_u$. Set $p=p+1$.
		 \label{step_3}
		\State  Update $\{s_{k}^{p+1} = s_{k}^\star$, $(\mathbf{m}_{k}^{e})^{p+1} = {(\mathbf{m}_{k}^{e})}^\star$, $\beta_{k}^{p+1}  =\beta_{k}^{\star}\}$ $\forall \hspace{2mm} k \in \mathcal U_e$  and  $\{(\mathbf{m}_{j}^{u})^{p+1} = {(\mathbf{m}_{j}^{u})}^\star$, $(b_{j}^{u})^{p+1}={(b_{j}^{u})}^\star\}$ $\forall \hspace{2mm} j \in \mathcal U_u$.
		 \label{step_4}
		
		\Statex \textbf{until} stopping criterion is satisfied
	\end{algorithmic}
\end{algorithm}

The algorithm is iterated until the difference between the objective values of problem \eqref{Optproblem3} in consecutive iterations is less than a predefined threshold.

\section{Numerical Results }
\label{results}
We simulate the proposed algorithm in order to prove the correctness and effectiveness of our algorithm. In our simulations, the downlink of a single-cell MISO system is considered. We assume that the base station is equipped with four transmit antennas. There are eight eMBB users and eight URLLC users in the system. To model the channel gains, we have used the exponential path loss model which is given by $\mathbf{h}_k={({r_k}/{r_0})}^{-\alpha}\mathbf{c}_k$, where $\mathbf{h}_k \in \mathbb C \tran$ is the channel vector from base station to $k$th user, $r_k$ is the distance from base station to $k$th user, $r_0$ is the far-field reference distance, $\alpha$ is the path loss exponent and $\mathbf{c}_k$ is small scale fading which is arbitrary chosen from circularly symmetric complex Gaussian vector distribution with mean zero and identity covariance matrix. We assume that both eMBB and URLLC users are distributed uniformly around the base station within the distance 10 m and 100 m. We consider the bandwidth allocation between eMBB users and URLLC users as total bandwidth for eMBB users, $b^e$ = $B_{total}\times ({1}/{2})$ and the total bandwidth of URLLC users, $b^u$ = $B_{total} \times ({1}/{2})$. Furthermore, the simulation parameters mentioned in Table \ref{table:simulation parameters} are assumed.

\begin{table}[ht]
	\caption{Simulation parameters} 
	\centering 
	\begin{tabular}{|c| c| } 
		
		\hline 
		Far field distance $r_0$ & 1 m \\
		\hline
		Path loss exponent $\alpha$  & 2 \\
		\hline
		Overall packet loss probability requirement $\epsilon$ & $1 \times 10^{-5}$  \\ 
		\hline
		Transmission error probability $\epsilon_c=\epsilon/2$& $5 \times 10^{-6}$  \\
		\hline
		Queueing-delay violation probability $\epsilon_q=\epsilon/2$& $5 \times 10^{-6}$  \\
		\hline
		E2E delay requirement $D_{max}$ & 1 ms  \\
	    \hline
		Maximum queueing delay $D_q$ & 0.8 ms  \\  
		\hline 
		Duration of each frame $T_f$ & 0.1 ms \\
		\hline 
		Duration of data transmission in one frame $\tau$ & 0.05 ms \\
		\hline
		Packet size $\mu$ & 20 bytes \\
		\hline
		Maximum transmit power $P_{total}$ & 33 dBm \\
		\hline
		Arrival packet rate $\lambda$ & 0.2 packets/frame \\
		\hline
		Single-sided noise spectral density $N_0$ & -83.98 dBm/Hz \cite{8761648}\\
		\hline
		Total bandwidth of the system $B_{total}$   & 200 MHz \\
		\hline
		Target rate for an eMBB user $R_{target}$  & 200 Mbps \\
		\hline
	\end{tabular}
	\label{table:simulation parameters} 
\end{table}

We simulate an arbitrarily chosen single channel and topology realization. The objective value $f(s) = {\sum}_{k=1 }^{K} \log(s_k + \delta)$ is calculated for every iteration until convergence. Furthermore, we count the admitted number of eMBB users at each iteration. Then we draw  the objective value versus iteration and number of admitted users versus iteration in the same graph in order to check the convergence of the algorithm.

\begin{figure}[ht]
\centerline{\includegraphics[width=0.5\textwidth]{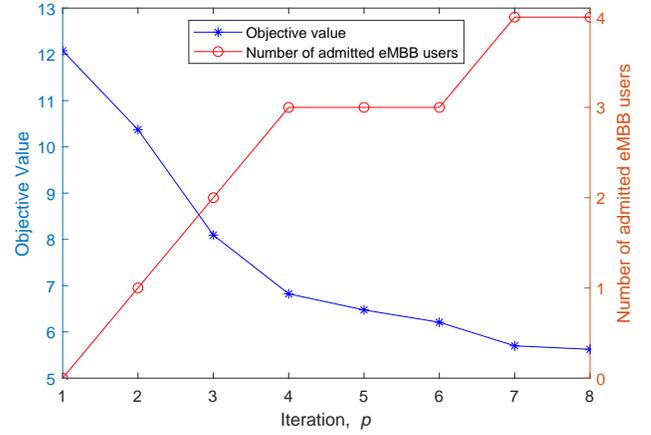}}
\caption{Objective value versus iterations and number of admitted users versus iterations.}
\label{fig:convergence}
\end{figure}

Fig. \ref{fig:convergence} shows the behaviour of the convergence of the \emph{Algorithm~\ref{algorithm 1}}. According to Fig. \ref{fig:convergence}, the objective value is minimized and converged after eight iterations. 
At the convergence we are able to get the optimal solution of the algorithm as four eMBB users can be admitted to the system. Therefore, we can observe that, our algorithm is obtaining the optimal solution in few iterations.

Next, we evaluate how the admitted number of eMBB users behave with the target rate for eMBB user and the total bandwidth of the system. The algorithm has been run over 100 channel and topology realizations. We simulate it for different ratios of eMBB and URLLC bandwidth allocation from the total bandwidth of the system. Table \ref{table:case} shows the two different cases that we have simulated. As a benchmark, we consider an exhaustive search algorithm. (We name \emph{Algorithm~\ref{algorithm 1}} as Algo 1).

\begin{table}[ht]
	\caption{eMBB and URLLC bandwidth allocation} 
	\centering 
	\begin{tabular}{|c| c |c| } 
	
		\hline 
		\bf{Case} & \bf{Bandwidth portion } & \bf{Bandwidth portion } \\
		&\bf{for eMBB} ($b^e$) & \bf{for URLLC} ($b^u$) \\
		
		\hline
		1 & $B_{total} \times ({3}/{4})$ & $B_{total} \times ({1}/{4})$ \\
		\hline
		2 & $B_{total} \times ({1}/{2})$ & $B_{total} \times ({1}/{2})$ \\
		\hline
	\end{tabular}
	\label{table:case} 
\end{table}

  \begin{figure}[htbp]
\centerline{\includegraphics[width=0.5\textwidth]{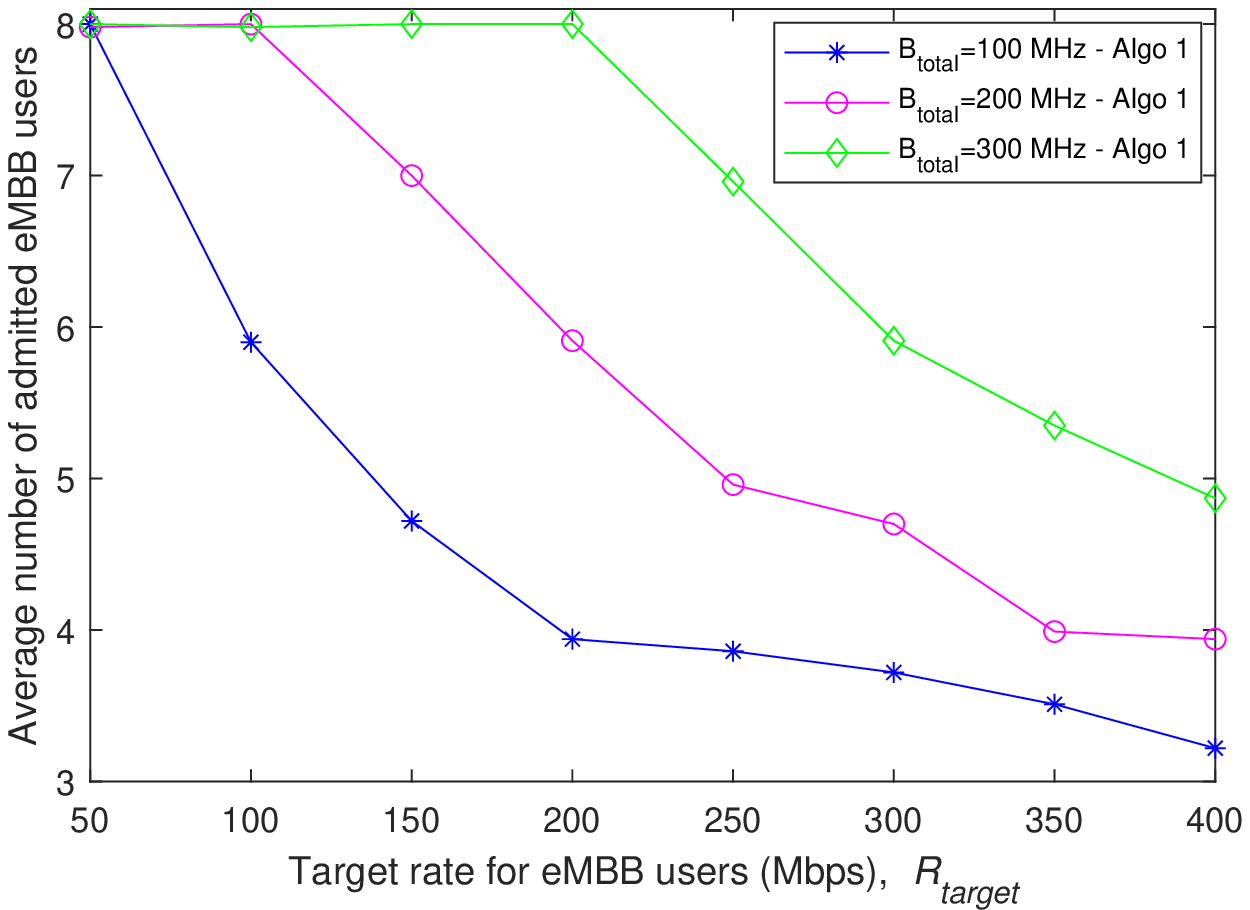}}
\caption{Admitted eMBB users versus target rate for eMBB users for case 1.}
\label{fig:vqrate}
\end{figure}

 \begin{figure}[htbp]
\centerline{\includegraphics[width=0.5\textwidth]{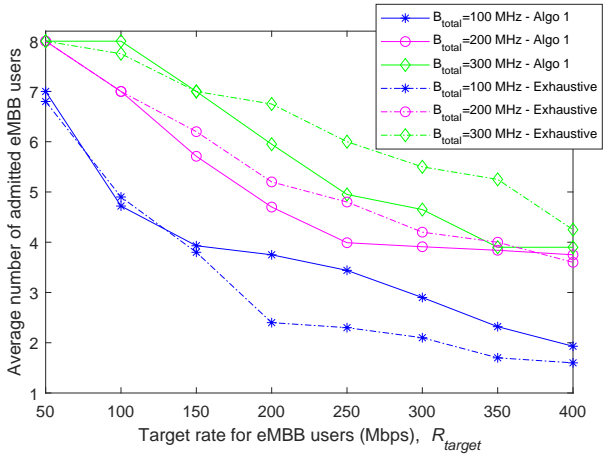}}
\caption{Admitted eMBB users versus target rate for eMBB users for case 2. }
\label{fig:vhrate}
\end{figure}

The variation of the admitted number of users with the target rate of eMBB users for different values of total bandwidth for case 1 and case 2 show in  Fig. \ref{fig:vqrate} and  \ref{fig:vhrate}  respectively.
 We can observe from Fig. \ref{fig:vhrate} that the average number of admitted users of \emph{Algorithm~\ref{algorithm 1}} is close to that of the exhaustive search algorithm. In addition, results show that when $R_{target}$ is high, our \emph{Algorithm~\ref{algorithm 1}} slightly outperforms the exhaustive search algorithm. Furthermore, from the Fig.  \ref{fig:vqrate} and \ref{fig:vhrate}, we can conclude that optimal bandwidth allocation between eMBB users and URLLC user is leading to a higher number of eMBB users admitted while satisfying the reliability and latency requirement of URLLC users. 

\begin{figure}[htbp]
\centerline{\includegraphics[width=0.5\textwidth]{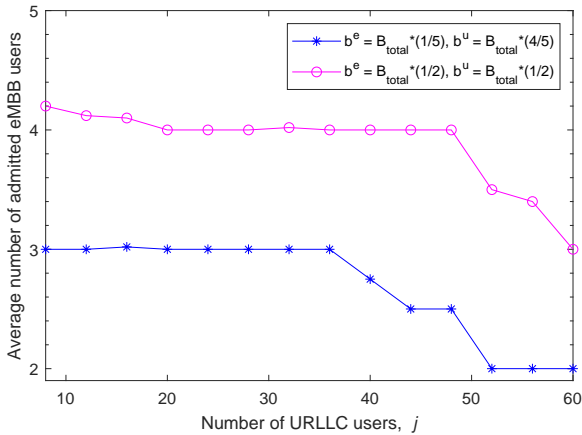}}
\caption{Admitted eMBB users versus number of URLLC users in the system.}
\label{fig:urllc_user}
\end{figure}

Further, we evaluate average number of admitted eMBB users versus number of URLLC users in Fig. \ref{fig:urllc_user}. 
Results show that even though the number of admitted eMBB users tends to decrease with the increase of URLLC users, optimal bandwidth allocation allows to have more eMBB users in the system while giving resources to the all URLLC users who have satisfied reliability and latency requirements. 

\section{Conclusion}
\label{conclusion}
We have considered the admission control problem in 5G networks where eMBB users and URLLC users are coexisting. URLLC users are time and mission critical while eMBB users need high data rates. Thus, for the coexistence of both URLLC and eMBB users, we need to control the admissions of eMBB users and facilitate scheduling all the URLLC users. 
In this paper we have proposed an admission control algorithm to maximize the number of admitted eMBB users to the system, who have sufficient data rate,  while allocating power, bandwidth and beamforming directions to all URLLC users whose latency and reliability requirements are always guaranteed. The proposed admission control algorithm is formulated as an $\ell_0$ minimization problem. It is based on $\ell_0$ approximation methods and sequential convex programming. Numerically we have shown that the proposed algorithm achieves near-optimal performance. From the numerical results, we can conclude that our system can maximize the possible number of admitted eMBB users with required data rate when all URLLC users have satisfied their reliability and latency requirements. Optimal bandwidth, power and beamforming directions allocation between eMBB users and URLLC user is leading to a higher number of eMBB users admitted. 
This research can be extended to address multi-cell scenario and user mobility.


\bibliographystyle{IEEEbib}

\bibliography{di}

\end{document}

%% file: commands.tex
\newcommand{\be}{\begin{equation}}
\newcommand{\ee}{\end{equation}}
\newcommand{\ba}{\begin{array}}
\newcommand{\ea}{\end{array}}
\newcommand{\bea}{\begin{eqnarray}}
\newcommand{\eea}{\end{eqnarray}}

\newcommand{\herm}{^{\mbox{\scriptsize H}}}  
\newcommand{\tran}{^{\mbox{\scriptsize T}}}  

\newcommand{\vbar}{\raisebox{.17ex}{\rule{.04em}{1.35ex}}}
\newcommand{\vbarind}{\raisebox{.01ex}{\rule{.04em}{1.1ex}}}
\newcommand{\R}{\ifmmode {\rm I}\hspace{-.2em}{\rm R} \else ${\rm I}\hspace{-.2em}{\rm R}$ \fi}
\newcommand{\T}{\ifmmode {\rm I}\hspace{-.2em}{\rm T} \else ${\rm I}\hspace{-.2em}{\rm T}$ \fi}
\newcommand{\N}{\ifmmode {\rm I}\hspace{-.2em}{\rm N} \else \mbox{${\rm I}\hspace{-.2em}{\rm N}$} \fi}
\newcommand{\B}{\ifmmode {\rm I}\hspace{-.2em}{\rm B} \else \mbox{${\rm I}\hspace{-.2em}{\rm B}$} \fi}
\newcommand{\Hil}{\ifmmode {\rm I}\hspace{-.2em}{\rm H} \else \mbox{${\rm I}\hspace{-.2em}{\rm H}$} \fi}
\newcommand{\C}{\ifmmode \hspace{.2em}\vbar\hspace{-.31em}{\rm C} \else \mbox{$\hspace{.2em}\vbar\hspace{-.31em}{\rm C}$} \fi}
\newcommand{\Cind}{\ifmmode \hspace{.2em}\vbarind\hspace{-.25em}{\rm C} \else \mbox{$\hspace{.2em}\vbarind\hspace{-.25em}{\rm C}$} \fi}
\newcommand{\Q}{\ifmmode \hspace{.2em}\vbar\hspace{-.31em}{\rm Q} \else \mbox{$\hspace{.2em}\vbar\hspace{-.31em}{\rm Q}$} \fi}
\newcommand{\Z}{\ifmmode {\rm Z}\hspace{-.28em}{\rm Z} \else ${\rm Z}\hspace{-.28em}{\rm Z}$ \fi}












%% file: main.bbl
\begin{thebibliography}{10}

\bibitem{8718159}
M.~{Alsenwi}, S.~R. {Pandey}, Y.~K. {Tun}, K.~T. {Kim}, and C.~S. {Hong},
\newblock ``{A Chance Constrained Based Formulation for Dynamic Multiplexing of
  eMBB-URLLC Traffics in 5G New Radio},''
\newblock in {\em 2019 International Conference on Information Networking
  (ICOIN)}, Jan 2019, pp. 108--113.

\bibitem{8241816}
K.~B. {Shashika Manosha}, S.~K. {Joshi}, M.~{Codreanu}, N.~{Rajatheva}, and
  M.~{Latva-aho},
\newblock ``{Admission Control Algorithms for QoS-Constrained Multicell MISO
  Downlink Systems},''
\newblock {\em IEEE Transactions on Wireless Communications}, vol. 17, no. 3,
  pp. 1982--1999, March 2018.

\bibitem{4570234}
E.~{Matskani}, N.~D. {Sidiropoulos}, Z.~{Luo}, and L.~{Tassiulas},
\newblock ``{Convex approximation techniques for joint multiuser downlink
  beamforming and admission control},''
\newblock {\em IEEE Transactions on Wireless Communications}, vol. 7, no. 7,
  pp. 2682--2693, July 2008.

\bibitem{5682069}
X.~{Gong}, S.~A. {Vorobyov}, and C.~{Tellambura},
\newblock ``{Joint Bandwidth and Power Allocation With Admission Control in
  Wireless Multi-User Networks With and Without Relaying},''
\newblock {\em IEEE Transactions on Signal Processing}, vol. 59, no. 4, pp.
  1801--1813, April 2011.

\bibitem{7945856}
C.~{She}, C.~{Yang}, and T.~Q.~S. {Quek},
\newblock ``{Radio Resource Management for Ultra-Reliable and Low-Latency
  Communications},''
\newblock {\em IEEE Communications Magazine}, vol. 55, no. 6, pp. 72--78, June
  2017.

\bibitem{8070468}
C.~{She}, C.~{Yang}, and T.~Q.~S. {Quek},
\newblock ``{Cross-Layer Optimization for Ultra-Reliable and Low-Latency Radio
  Access Networks},''
\newblock {\em IEEE Transactions on Wireless Communications}, vol. 17, no. 1,
  pp. 127--141, Jan 2018.

\bibitem{8541123}
C.~{Sun}, C.~{She}, C.~{Yang}, T.~Q.~S. {Quek}, Y.~{Li}, and B.~{Vucetic},
\newblock ``{Optimizing Resource Allocation in the Short Blocklength Regime for
  Ultra-Reliable and Low-Latency Communications},''
\newblock {\em IEEE Transactions on Wireless Communications}, vol. 18, no. 1,
  pp. 402--415, Jan 2019.

\bibitem{8269133}
C.~{Sun}, C.~{She}, and C.~{Yang},
\newblock ``{Exploiting Multi-User Diversity for Ultra-Reliable and Low-Latency
  Communications},''
\newblock in {\em 2017 IEEE Globecom Workshops (GC Wkshps)}, Dec 2017, pp.
  1--6.

\bibitem{8408793}
A.~A. {Esswie} and K.~I. {Pedersen},
\newblock ``{Opportunistic Spatial Preemptive Scheduling for URLLC and eMBB
  Coexistence in Multi-User 5G Networks},''
\newblock {\em IEEE Access}, vol. 6, pp. 38451--38463, 2018.

\bibitem{8746407}
A.~{Karimi}, K.~I. {Pedersen}, N.~H. {Mahmood}, G.~{Pocovi}, and P.~{Mogensen},
\newblock ``{Efficient Low Complexity Packet Scheduling Algorithm for Mixed
  URLLC and eMBB Traffic in 5G},''
\newblock in {\em 2019 IEEE 89th Vehicular Technology Conference
  (VTC2019-Spring)}, April 2019, pp. 1--6.

\bibitem{8761648}
J.~{Tang}, B.~{Shim}, T.~{Chang}, and T.~Q.~S. {Quek},
\newblock ``{Incorporating URLLC and Multicast eMBB in Sliced Cloud Radio
  Access Network},''
\newblock in {\em ICC 2019 - 2019 IEEE International Conference on
  Communications (ICC)}, May 2019, pp. 1--7.

\bibitem{thesis}
Nipuni~Uthpala Ginige,
\newblock ``{ADMISSION CONTROL IN 5G NETWORKS FOR THE COEXISTENCE OF EMBB-URLLC
  USERS},''
\newblock {\em URL:
  \url{http://jultika.oulu.fi/Record/nbnfioulu-201910223007}}, 2019.

\bibitem{2019admission}
Nipuni~Uthpala Ginige, KB~Manosha, Nandana Rajatheva, and Matti Latva-aho,
\newblock ``{Admission Control in 5G Networks for the Coexistence of eMBB-URLLC
  Users},''
\newblock {\em arXiv preprint arXiv:1910.13855}, 2019.

\bibitem{candes2008enhancing}
Emmanuel~J Candes, Michael~B Wakin, and Stephen~P Boyd,
\newblock ``{Enhancing sparsity by reweighted $\ell_1$ minimization},''
\newblock {\em Journal of Fourier analysis and applications}, vol. 14, no. 5-6,
  pp. 877--905, 2008.

\bibitem{scp}
S.~Boyd,
\newblock ``{Sequential Convex Programming},''
\newblock [Online]. Available:
  \url{https://stanford.edu/class/ee364b/lectures/seq_slides.pdf}.

\bibitem{boyd2004convex}
Stephen Boyd and Lieven Vandenberghe,
\newblock {\em {Convex optimization}},
\newblock Cambridge university press, 2004.

\end{thebibliography}
